\documentclass{aa}
\usepackage{graphicx}
\usepackage{longtable}
\usepackage{psfig}
\usepackage{amsmath} 
\usepackage{aas_macros} 
\usepackage{natbib}  
\bibpunct{(}{)}{;}{a}{,}{,}

\def\r{~X-ray rotational~}
\def\m{~modulation~}
\begin{document}

\title{X-ray rotational modulation of a supersaturated star in IC 2391} 
\author{A. Marino \inst{1}  
\and
G. Micela\inst{2}
\and
G. Peres\inst{1}
\and
S. Sciortino\inst{2}
}

\institute{Dipartimento di Scienze Fisiche e Astronomiche,
Sez. di Astronomia, Universit\`a di Palermo,
Piazza del Parlamento 1, I-90134 Palermo, Italy\\
\and
INAF - Osservatorio Astronomico di Palermo G.S. Vaiana,
Piazza del Parlamento 1, I-90134 Palermo, Italy\\
}

\offprints{marino@astropa.unipa.it}

\date{Received xxx; accepted  xxx}

   \abstract{We present evidence of  X-ray rotational modulation  
on VXR45, a young fast rotator star, member of IC 2391.  
It is a dG9 spectral type star whose rotational period and 
X-ray luminosity  make it a supersaturated star.
Our X-ray observation, made with EPIC/PN on XMM-Newton, covers about
two photometric rotational periods.
The detection of \r \m implies the presence of  structural inhomogeneities. 
Possible interpretations are presented and discussed.
}

\titlerunning{X-ray rotational modulation of a supersaturated star}
\authorrunning{Marino et al.} 
\maketitle

\section{Introduction\label{intro}}

Rotational modulation by a factor of two, due to the non-uniform distribution
of emitting regions, is observed in solar coronal emission
at any phase of the solar cycle. 
Since we cannot resolve stellar coronal structures, rotational modulation
is one of the main methods to find evidence of non-uniform distribution of 
emitting stellar coronal regions  and to investigate geometrical and 
physical properties of stellar coronal structures.
Therefore, detection of \r \m  on a very active star 
is particularly important to gain some insight into the physical mechanisms 
which operate on these stars.
Observational data have shown that the most  active stars, rotating above
$\sim$ 15-20 km/s, reach a maximum X-ray luminosity such that 
L$_{{\rm x}}$/L$_{{\rm bol}}$ saturates at $\sim$ 10$^{-3}$, 
where L$_{\rm bol}$ indicates the star's bolometric luminosity \citep[e.g.][]{Vi84}.  
Saturation lacks a clear  interpretation: it could be the effect of dynamo saturation,
or rather could correspond to the total filling  of the star's surface
by active regions, as originally suggested by \citet{Vi84}.
Furthermore ROSAT observations  have shown that very fast rotators ($ v \sin i >$ 100
km/s) exhibit a level of X-ray luminosity a factor of 3-5 below the saturated level
\citep{Ra98}. 
The origin of this phenomenon, named supersaturation by \citet{Pro96} 
is still unexplained and several hypotheses have been formulated.
Supersaturation could be the result of an overall decrease of dynamo efficiency 
at very high rotation rate or the consequence of the
redistribution of the radiative output due to the lower stellar gravity because
of the enhanced rotation \citep{Ra98}.
A different explanation is that very rapid rotation could
lead to a higher coronal temperature and to the shift of the X-ray emission out of the 
ROSAT passband \citep{Ra98}. 
\citet{St01} suggested a common explanation for the supersaturation both in ultra fast
rotators stars (UFRs) and in W UMa binaries via a decreased coverage of 
the stellar surface by X-ray emitting regions, although 
they noted that the physical mechanisms leading to such a decreased coverage 
can be different for the  two star classes. For UFRs they proposed strong 
polar updrafts within a convection zone, driven by nonuniform heating from below.
Thanks to the high sensitivity and broad bandpass of XMM-Newton,
we can now test some of these hypotheses. However presently there is no 
consensus on phenomena occurring in a stellar corona either in the saturated or 
in supersaturated regime. 

There have been several attempts  to observe  rotational
modulation of stellar coronal emission in fast rotating 
stars \citep[e.g.][]
{Col88, Vi92, AgV88, ku92, D94, Gu95, ku97, Au01, Ga03} in order to find 
a possible signature of non-uniform
distribution of emitting coronal regions. 
In most cases the results were affected by frequent 
superimposed flares or poor sampling over more than one rotation period
\citep[e.g.][]{Col88,Vi92}. 
Probably one of the  best examples  of  \r \m is that reported by
\citet{Gu95}: X-ray flux variations within a factor of two in the
ROSAT All-Sky Survey light-curve of  EK Draconis, a dG0 star with an optical
rotation period of 2.7-2.8 day, just slightly too long for the star to be in the
saturation regime.
\citet{Gu95} found that the X-ray light curve is significantly variable in the
ROSAT energy  band with the emission from the cooler plasma being
responsible for the rotational \m but with the hot plasma approximately
constant. 
On the other hand  \citet{ku97} find only a partial 
evidence of \r \m of the X-ray flux  implying structural inhomogeneities 
in the saturated star AB Dor. Similar results are found in BeppoSAX observations 
of AB Dor by \citet{Fra02}.
On the contrary \citet{Gu01} find no evidence for \r modulation in a  
XMM-Newton observation of  AB Dor.
Analogously also \citet{Si99} find no indication of \r \m both in the saturated fast 
rotators HD197890, of similar spectral type of VXR45, and in Gl890,
observed with ASCA.

In this {\em Letter} we present the unambiguous  detection of  \r
\m in a fast rotating star member of IC 2391: VXR45, observed with XMM-Newton/EPIC/PN.
VXR45 is a dG9 star with a very short photometric period of 0.223 days
\citep{Pa96}, and  $ v \sin i >$ 200 km/s \citep{Stau97}. Its logL$_{\rm x}$, 
as measured with ROSAT, ranges between 30.12 and 30.10 erg/s, but the range of 
log(L$_{\rm x}$/L$_{\rm bol}$) between $-$3.60 and $-$3.62 puts 
the star below the saturation level,  making it a supersaturated star.

\section{The observation and data analysis}
The observation analyzed here was dedicated to the young open cluster IC 2391 
as part of  EPIC/PN GTO time and it  has been performed in November 20, 2001. 
Here we present  EPIC/PN data relative to the star VXR45, 
located at RA (J2000) = 8$^{h}$ 42$^{m}$ 14.8$^{s}$, Dec
(J2000) = $-$52$^{o}$ 56$^{'}$ 01$^{''}$; it has  V=10.70 and
B$-$V=0.81. 

The nominal duration of the observation is 50 ks; after the selection of the good time
intervals (i.e. those of adequately low background emission) we retained $\sim$ 33 ks. 
The X-ray data reduction was performed with the {\em SAS} 5.4.1 software
\footnote{available at http://xmm.vilspa.esa.es/sas}.
The star is one of the brightest  in the field and it is located 
3.96$^{'}$ off-axis, far from the EPIC/PN  CCD gaps.

\subsection{Light-curve}
\begin{figure}[h]
\centerline{\psfig{figure=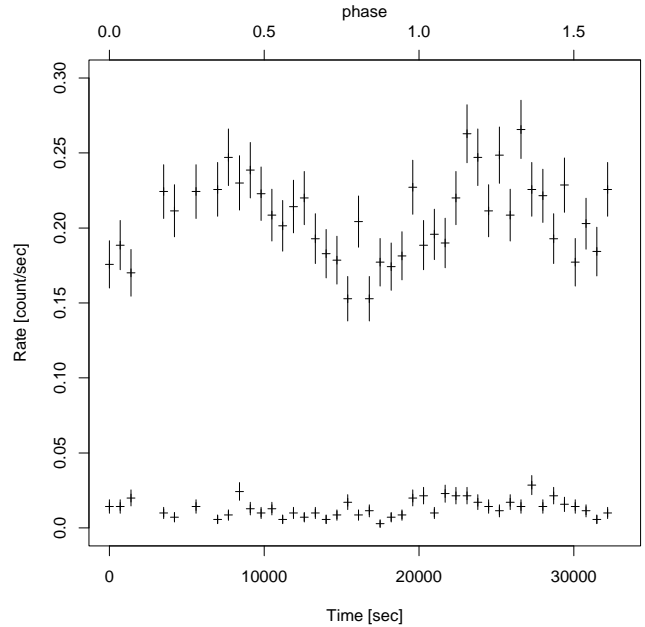,width=9.0cm}}
\caption{X-ray light-curve of  VXR45 (top) and background (bottom)  as seen with EPIC/PN in the 0.3 - 7.8 keV band, time bins are 700 sec long. The photometric rotational period of $\sim$ 19.3 ksec  is well visible.}
\label{Ltc}
\end{figure}

In order to determine the count rate, we extracted photons from a circular
region centered on the source and selected so as to include $\sim$ 80\% 
of the source photons \citep{Ghi01,Saxon02}, corresponding to  a radius of 37$^{''}$. 
The total number of photon counts in the 0.3 - 7.8 keV band  
amounts to 7431; 
the local background has been determined from a source-free region near VXR45 
for a total of  506 counts.
Fig. \ref{Ltc} shows the light-curves of VXR45 obtained in the 0.3 - 7.8 keV band
(top) and of  the corresponding background (bottom).
The X-ray  modulation, with a period very similar to the photometric 
one (0.223 days = 19.3 ksec), 
is clearly evident in Fig. \ref{Ltc}. Note that other short term  variability 
not due to \r \m  may also be present in the light-curve 
but the data clearly  show  that the variability observed cannot be explained 
with individual large flares. 

\begin{figure}[h]
\centerline{\psfig{figure=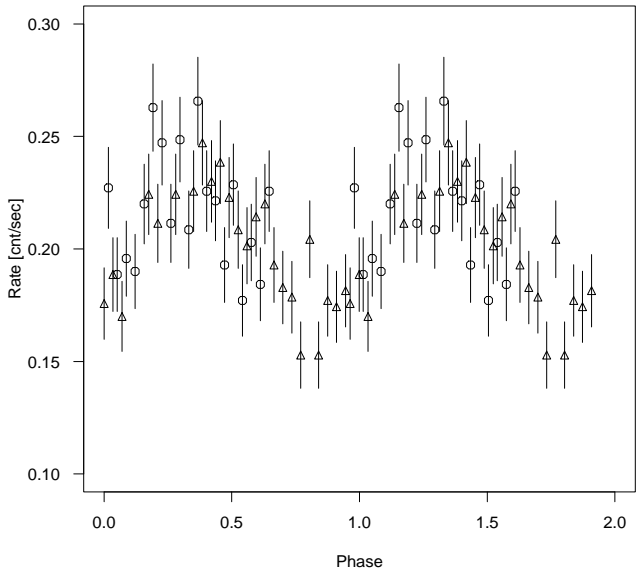,width=9.0cm}}
\caption{X-ray data of Fig. \ref{Ltc} folded with the rotational period vs.
 phase. Circles are points observed  at t $<$ 0.223 days (the
photometric period) since the observation start and triangles those 
observed at t $>$ 0.223 days. For clarity we report two photometric periods.}
\label{fo}
\end{figure}

\begin{figure}[h]
\centerline{\psfig{figure=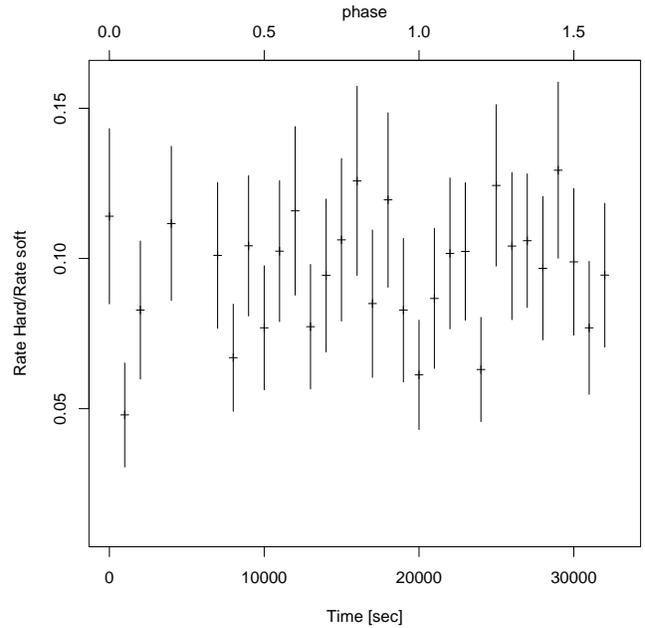,width=9.0cm}}
\caption{Ratio of hard count rate (1.5 - 3.5 keV) to soft count rate (0.3 - 1.5 keV) vs. time for VXR45, time bins are 1000 sec long. There is no evidence for modulation of the ratio.}
\label{LSH}
\end{figure}

In order to confirm that modulation of the X-ray emission is due to rotation we
have folded the X-ray light-curve  with the photometric rotational period 
as shown in  Fig. \ref{fo}. Phase-related variability is clearly evident; 
the X-ray variations in the light-curve are $\sim$ 30\% of the average count rate.
Using a radius of 0.9 R$_{\odot}$ adequate for the star's age and spectral type
\citep{Sie00} and knowing P$_{\rm rot}$ and $ v \sin i$, we obtain an inclination 
angle close to 90$^o$. This makes the star an ideal target  to detect
rotational modulation due to a longitudinal structure.
In order to explore if differences exist between modulation of cool and hot plasma
as in EK Dra {\citep{Gu95}} we show in Fig. \ref{LSH} the ratio between the 
hard count rate (1.5 - 3.5 keV) and the soft count rate (0.3 - 1.5 keV) versus 
the time \footnote{The break point 1.5 keV separates the line-dominated 
Fe L shell + {\em NeIX/X + MgXI/XII} region from the part
dominated by bremsstrahlung continuum. Hence the adopted ratio maps on temperature.}. 
There is no strong evidence of spectral changes in the soft and hard passbands,
and the $\chi^2$ test applied to the hard/soft count-rate ratio confirms that the
variations observed are those expected in a random sample.
We tried different energy boundaries in soft and hard bands definition and the
result does not change.
The lack of modulation in  the hard (1.5 - 3.5 keV)/soft (0.3 - 1.5 keV) ratio 
is consistent with the hypothesis that hotter and cooler plasmas are co-spatial. 
On the contrary the \r \m of EK Dra \citep{Gu95} was detected predominantly in 
the soft ROSAT band (0.1 - 0.4 keV), that in practice is outside the XMM-Newton 
bandpass, while it was approximately constant in the hard ROSAT band (0.4 - 2 keV).

\subsection{Count Rate, Flux and Luminosity}
\begin{figure}[h]
\centerline{\psfig{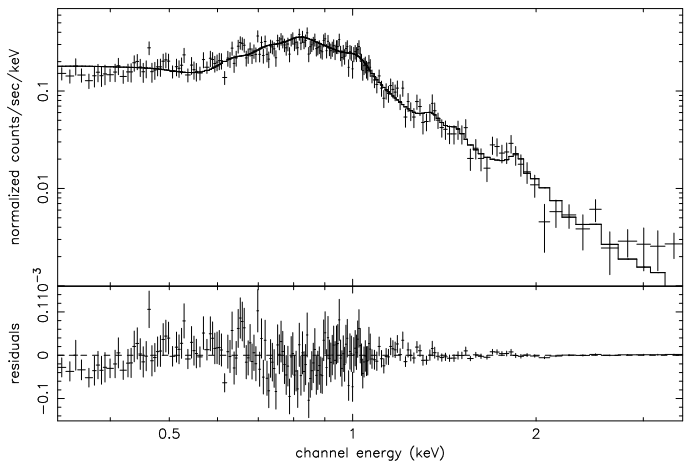}}
\caption{X-ray spectrum of VXR45 with the best fit model superimposed.} 
\label{Spe}
\end{figure}

We extracted the spectrum of the single and doubly flagged PN events with energy
in 0.3 - 3.5 keV band that contains the essential part of the source (Fig. \ref{Spe}).
We found that a 2-T APEC spectral model provides the best fit to the data,
as measured with a $\chi^2$ test ($\chi_{\nu}^2$=1.01 for 189 degrees of freedom).
The N$_H$ value was fixed to 10$^{19.5}$ cm$^{-2}$ \citep{Pa96}.
The two temperatures in the model, as well as the corresponding emission measures
and the abundance Z were allowed to vary as free parameters in the fitting 
process. We obtained temperatures of 0.61 keV and 1.19 keV, with a ratio 
of emission measure of EM$_{\rm cold}$/EM$_{\rm hot} \approx$ 1.42, and 
Z $\sim$ 0.27Z$_\odot$. 
The spectral fit yielded an  X-ray flux, in the 0.3 - 3.5 keV band, 
of  4.77 $\times$ 10$^{-13}$ ergs~cm$^{-2}$~s$^{-1}$ corresponding 
to an  average X-ray luminosity of 1.50~$\times$~10$^{30}$ erg/s  adopting a 
distance of 162 pc \citep{Pa96} and log(L$_{\rm x}$/L$_{\rm bol}$) =$-$3.54. 
Assuming that the spectrum does not change, as suggested by Fig. \ref{LSH}, the  peak 
to peak variations  of log(L$_{\rm x}$) are 30.04 - 30.28 and log(L$_{\rm x}$/L$_{\rm bol}$)
ranges between $-$3.44 and $-$3.68. These intervals include the values observed 
with ROSAT.

\section{Discussion and conclusions}
Notwithstanding the wealth of available data, \r \m is very hard to detect 
in active stars, as discussed in the Introduction.
We observed VXR45 a fast rotator member of IC 2391, with XMM-Newton/EPIC/PN
for $\sim$  33 ks, corresponding to $\sim$ 2 photometric rotational periods. 
We find unambiguous evidence of  \r \m  from a supersaturated  star
with a level of X-ray emission $\sim$ 100 times larger than the solar one. 
We note that  we detect \r \m in the  0.3 - 7.8  keV energy band  while the
\r \m of EK Dra \citep{Gu95}, was detected in the softer and smaller  ROSAT
energy band (0.1 - 0.4 keV); furthermore VXR45 is a dG9 supersaturated star 
located on the left of the logL$_{\rm x}$/L$_{\rm bol}$ vs. P$_{\rm rot}$ plot 
(being a very fast rotator) \citep{Ra98} while  EK Dra is
a dG0 with a similar log(L$_{\rm x}$/L$_{\rm bol}$)=$-$3.53  
but a rotation period 10 times slower that puts it on the right side
of the same plot and just before the saturated regime \footnote{L$_{\rm x}$/L$_{\rm
bol}$ has been obtained taking L$_{\rm x}$ = 10$^{29.92}$ erg/s \citep{Gu95} and
L$_{\rm bol}$ = 2.70 $\times$ 10$^{33}$ erg/s computed using the transformations
by \citet{Flo96}.}.
On the other hand \r \m has been detected  in at least some observations in AB Dor,
a star in the saturated regime, while it has not been detected in other saturated
stars (HD 197890, Gl890).

VXR45 was extensively observed with ROSAT, but an analysis of  the PSPC 
observations of those IC 2391 cluster members whose  photometric rotational 
period was known (including VXR45), failed to detect  \r \m in any of the X-ray 
light-curves \citep{Pa96}. 
These authors  note that the  segments of their observations were not optimally spaced
for this kind of analysis.
Furthermore, they studied the time variations of the PSPC VXR45 light-curve, with
several statistical tests but obtaining  contradicting  results.

The new  XMM-Newton observation allows us to detect \r \m thanks to the
quasi-continuous time coverage, spanning twice the photometric period.
The detection of \r modulation, implying the presence of non-uniformly distributed
active regions on the star, is in some sense surprising, since  bright coronae 
variations are typically due to flares.

Thanks to the broad energy bandpass of XMM-Newton, we can exclude the 
hypothesis that this supersaturated star has higher coronal 
temperature that may cause most of the X-ray emission to be outside the ROSAT passband.
The detection of \r \m indicates that the star is not completely covered by
active regions. Furthermore, as shown in Fig. \ref{LSH} there is no strong
evidence for spectral variations in the  soft and hard passbands. 
This is consistent with  the hypothesis that the modulation we observe is 
mainly due to a longitudinal concentration of X-ray emitting material, 
and that at all times the emission is largely due to the same mixture of 
emitting structures.
The average surface X-ray flux, derived from the  X-ray luminosity divided by the 
stellar surface is  3.0 $\times$ 10$^7$ erg~s$^{-1}$~cm$^{-2}$ 
and  ranges between 3.9 $\times$ 10$^7$  and 2.2 $\times$ 10$^7$ erg~s$^{-1}$~cm$^{-2}$.
These surface X-ray fluxes values are a bit higher than those of solar active regions 
\citep{Or01} but intermediate between active regions and solar flares \citep{Re01},
and the X-ray temperatures are also intermediate between solar active regions  
and the flares. 

Our finding of a incomplete coverage of supersaturated stellar surface
are qualitatively in agreement with the explanation proposed by \citet{St01}.
A possible scenario is that a large polar region is always present and  another 
region at lower latitude is responsible for the observed modulation, 
albeit other scenarios are possible.
Observations of other saturated and supersaturated stars will help in
clarifying this picture.

\begin{acknowledgements}{We acknowledge financial support from ASI and MIUR. We wish
to thank the referee  M. Guedel for useful comments.}
\end{acknowledgements}
\bibliographystyle{aa}
\bibliography{Fd074}
\end{document}